\documentclass[aps,prd,twocolumn,showpacs,10pt,superscriptaddress,preprintnumbers,nofootinbib]{revtex4-2}

\usepackage{hhline}
\usepackage{stackengine}
\usepackage{subcaption}
\usepackage{vmargin,epsfig}
\usepackage{graphicx}
\usepackage[english]{babel}
\usepackage[toc,page]{appendix}
\usepackage{parskip}
\usepackage{xspace}
\usepackage{braket}
\usepackage{mathtools}
\usepackage{multirow}
\usepackage{booktabs}
\usepackage{subcaption}
\usepackage[colorlinks=true]{hyperref}
\hypersetup{
colorlinks=true,
linkcolor=blue,
linktocpage=true,
citecolor=violet,
urlcolor=blue
}
\allowdisplaybreaks

\usepackage{etoolbox}
\makeatletter
    \newcommand\newsubsupcommand[4]{\newcommand#1{#2\sc@subp{#3}{#4}}}
    \def\sc@subp#1#2{%
        \let\sc@subflag\undefinded%
        \let\sc@supflag\undefinded%
        \def\sc@thesub{#1}%
        \def\sc@thesup{#2}%
        \sc@proc%
    }%
    \def\sc@proc{%
        \@ifnextchar{_}{\def\sc@subflag{}\sc@mergesubs}{%
            \@ifnextchar{^}{\def\sc@supflag{}\sc@mergesups}{
                \ifdef{\sc@subflag}{}{_{\sc@thesub}}%
                \ifdef{\sc@supflag}{}{^{\sc@thesup}}%
            }%
        }%
    }%
    \def\sc@mergesubs#1#2{_{\sc@thesub#2}\sc@proc}%
    \def\sc@mergesups#1#2{^{\sc@thesup#2}\sc@proc}%
\makeatother

\newsubsupcommand{\phiL}{\varphi}{L}{}
\newsubsupcommand{\phiR}{\varphi}{R}{}
\newsubsupcommand{\psiL}{\psi}{L}{}
\newsubsupcommand{\psiR}{\psi}{R}{}
\newcommand{\namedref}[2]{\hyperref[#2]{#1~\ref*{#2}}}

\newcommand{\FxsQED}{$F_\pi\!\times\!\mathrm{sQED}$\xspace}
\makeatletter
\def\mr@ignsp#1 {\ifx\:#1\@empty\else #1\expandafter\mr@ignsp\fi}%
\newcommand{\multiref}[1]{\begingroup%\let\protect\string%
\xdef\mr@no@sparg{\expandafter\mr@ignsp#1 \: }%
\def\mr@comma{}%
\@for\mr@refs:=\mr@no@sparg\do{\mr@comma\def\mr@comma{,\,}\ref{\mr@refs}}%
\endgroup}
\makeatother
\newcommand{\liverpool}{Department of Mathematical Sciences, University of Liverpool, Liverpool L69 3BX, 
U.K.
}
\newcommand{\kharkov}{National Science Center KIPT, Kharkov, Ukraine
}

\usepackage{orcidlink}

\usepackage{tikz}
\usetikzlibrary{"arrows", "automata", "backgrounds", "calendar", "chains", "matrix", "mindmap", "patterns", "petri", "shadows", "shapes.geometric", "shapes.misc", "spy", "trees"}
\usetikzlibrary{arrows,shapes}
\usetikzlibrary{trees}
\usetikzlibrary{matrix,arrows} 	
\usetikzlibrary{positioning}				% For "above of=" commands
\usetikzlibrary{calc,through}				% For coordinates
\usetikzlibrary{decorations.pathreplacing}
\usepackage{pgffor}
\usepackage{mathdots}

\usetikzlibrary{decorations.pathmorphing}	% For Feynman Diagrams
\usetikzlibrary{calc}
\usetikzlibrary{decorations.markings}
\tikzset{
    fermion/.style={draw=black, postaction={decorate},
        decoration={markings,mark=at position .55 with {\arrow[draw=black]{>}}}},
    photon/.style={decorate, draw=black,
        decoration={coil, aspect=0,amplitude=1.5pt, segment length=4.5pt}},        
    scalar/.style={dashed,draw=black, postaction={decorate}}
}

\definecolor{green1}{HTML}{244819}
\definecolor{cyan1}{HTML}{37cdaa}
\definecolor{blue1}{HTML}{5d7ac4}
\definecolor{red1}{HTML}{d0482a}
\definecolor{purple1}{HTML}{845ea8}
\definecolor{orange1}{HTML}{e07229}

\begin{document}
\preprint{}

\title{Radiative return meets GVMD}

\author{Pau Petit~Ros\`as\orcidlink{0009-0009-8824-5208}\,}\email{paupetit@liverpool.ac.uk}
\affiliation{\liverpool} 
\author{Olga~Shekhovtsova\orcidlink{0000-0002-4237-8170}\,}\email{shekhovt@lnf.infn.it}
\affiliation{\kharkov} 
\author{William J. Torres~Bobadilla\orcidlink{0000-0001-6797-7607}\,}\email{torres@liverpool.ac.uk}
\affiliation{\liverpool} 
%

%\date{}

\begin{abstract}
We improve the description of pion-photon interactions in the radiative return process $e^+e^-\to \pi^+\pi^-\gamma$ at the next-to-leading order by including the pion form factor in the Feynman rules. We present a general calculation of the new amplitudes, and provide an implementation easy to interface with any Monte Carlo generator. We incorporate this framework into the event generator \texttt{Phokhara} and study several experimental configurations. Overall, we find percent-level effects appearing in angular differential cross section distributions at colliders whose centre-of-mass energies lie near the peak of the pion form factor. By contrast, total cross sections and distributions in charge-even variables show effects only at the permille level, or no visible differences at all. Finally, we compare the new predictions with KLOE measurements of the forward–backward asymmetry in order to assess the predictive power of the modifications.
\end{abstract}

\maketitle

\section{Introduction}
The muon anomalous magnetic moment, \(a_\mu~=~(g~-~2)_\mu/2\) has been measured with a remarkable relative precision of \(127\)\,ppb~\cite{Muong-2:2025xyk, Muong-2:2023cdq, Muong-2:2021ojo}, by the dedicated Fermilab \(g-2\) experiment and its predecessor at Brookhaven National Laboratory. Owing to its sensitivity to quantum effects, \(a_\mu\) constitutes a stringent test of the Standard Model, and the comparison between theory and experiment has shown a tension for more than a decade~\cite{Aoyama:2020ynm,Blum:2013xva}. In this context, the hadronic vacuum polarization contributions (HVP) are of particular relevance: they provide a numerically significant contribution and remain a leading source of theoretical uncertainty. Since they are governed by non-perturbative QCD at low energies, they must be obtained either from first principles using lattice QCD, the choice in WP25~\cite{Aliberti:2025beg} for the $a_\mu$ theoretical average, through dedicated experiments such as MUonE~\cite{CarloniCalame:2015obs,MUonE:2016hru,Banerjee:2020tdt}, or indirectly through measurements of the \(e^+e^- \to \text{hadrons}\) cross section at GeV-scale electron--positron colliders~\cite{Keshavarzi:2018mgv,Keshavarzi:2024wow,Davier:2019can}. Recent lattice QCD results~\cite{Borsanyi:2020mff,RBC:2024fic,ExtendedTwistedMass:2024nyi,Bazavov:2024eou} and the energy-scan measurement from the CMD-3 collaboration~\cite{CMD-3:2023alj,CMD-3:2023rfe} suggest a total alleviation of the long-standing discrepancy, reducing the previously quoted $\sim 5\sigma$ tension to a $1-2\sigma$ agreement. It is therefore crucial to understand the origin of the differences between lattice QCD determinations and those obtained with experiment-based methods, as well as the reason behind the difference between the most recent CMD-3 measurement and previous data driven results.

A central role in this discussion is played by pion-pair production, \(e^+e^- \to \pi^+\pi^-\), which accounts for more than \(70\%\) of the HVP contribution to \(a_\mu\). The experimental program for this channel is currently active: new radiative return studies---characterized by the radiative process \(e^+e^- \to \pi^+\pi^-\gamma\) to scan the pair production cross section at different energies---by BaBar~\cite{BaBar:2012bdw}, Belle II~\cite{Mori:2007bu}, BESIII~\cite{BESIII:2015equ}, and KLOE~\cite{KLOE:2008fmq, KLOE:2010qei, KLOE:2012anl, KLOE-2:2017fda} are underway, and updated energy-scan results, which rely on measuring the pion pair production channel at different centre of mass energies, for the CMD-3 and SND collaborations are anticipated. The \(\pi^+\pi^-\) channel is also the focus of the recent CMD-3 analysis, which introduced, among other elements, a dedicated investigation of the forward--backward asymmetry~\cite{Ignatov:2022iou, Colangelo:2022lzg}. These studies underscored the importance of pion compositeness effects, which had been neglected in earlier treatments. In particular, predictions based on the standard \FxsQED treatment, effectively describing pions as point-like scalar particles, governed by scalar QED (sQED), dressed by a vector form factor (VFF), were found to strongly disagree with the CMD-3 forward--backward asymmetry data.

This discrepancy was resolved by approaches that incorporate the pion vector form factor, $F_\pi$, consistently within loop corrections. The first of these was a framework inspired by generalized vector-meson dominance (GVMD), introduced in~\cite{Ignatov:2022iou}, in which $F_\pi$ is fitted by a sum of Breit--Wigner functions. Soon after, the FsQED model~\cite{Colangelo:2014dfa,Colangelo:2015ama,Colangelo:2022lzg} was developed, relying instead on dispersive relations to determine $F_\pi$.
%\footnote{It is worth mentioning GVMD approach corresponds to the pion-pole contribution in the \FxsQED model, see Eq.~(70) in~\cite{Fang:2025mhn}.} 
In both cases, the form factor enters the loop amplitudes in a well-defined manner, thereby eliminating the ambiguity inherent to \FxsQED regarding the momentum at which \(F_\pi\) should be evaluated. These improved descriptions of the energy-scan channel have been implemented in several Monte Carlo~(MC) generators: \texttt{MCGPJ}~\cite{Arbuzov:2005pt}, modified to include GVMD, \texttt{Babayaga@NLO}~\cite{Balossini:2006wc,Budassi:2026lmr}, extended to include both GVMD and FsQED~\cite{Budassi:2024whw}, and \texttt{McMule}~\cite{Banerjee:2020rww}, which employs an effective field theory formulation to realize FsQED~\cite{Fang:2025mhn}. 

Two comments are in order. First, it is worth noting that the GVMD approach employs a parametrization of the pion vector form factor that does not respect analyticity constraints, as it is based on a sum of Breit--Wigner functions. This has no visible impact on the energy scan mode~\cite{Budassi:2024whw}. 

Second, similar studies have not been done for the radiative return method. In fact, several open issues remain in this channel. They ultimately trace back to the fact that, with current technology, the full Compton tensor cannot be embedded in loop calculations. Furthermore, for the radiative return, it is not known if FsQED or GVMD provide an accurate description of the one-loop contributions which contain \(\gamma^*\gamma^*\to\pi^+\pi^-\gamma\)~\cite{Aliberti:2024fpq}. Nevertheless, implementing these models for the other diagrams is already a meaningful step that directly improves the theoretical description, while an implementation for all cases might be useful in the future.

We bridge the gap between the energy-scan and radiative return approaches by providing a calculation of the GVMD modifications in a form that is straightforward to interface with any MC generator, providing a \texttt{Fortran} implementation in the Ancillary files~\cite{zenodo}. Then, to assess the impact of the GVMD corrections in the radiative channel, we implement the code in the widely used MC generator \texttt{Phokhara}~\cite{Rodrigo:2001jr,Rodrigo:2001kf,Kuhn:2002xg,Czyz:2002np,Czyz:2003ue,Czyz:2017veo,Campanario:2019mjh}, which has been the tool of choice at \(\phi\)- and \(B\)-factories for radiative return studies. Finally, we use the new version of \texttt{Phokhara} to measure the impact of the GVMD modifications in a set of benchmark scenarios, defined in~\cite{Aliberti:2024fpq}, as well as to compare the MC predictions with KLOE data for the forward--backward asymmetry. In the latter, we remark the importance of additional contributions %described in Appendix~\ref{Appendix:ISC} 
coming from meson resonances.
~

This paper is organized as follows. 
%
% In Section~\ref{SecI}, we make a brief overview of the GVMD model and present the implementation strategy and validation of the corrections for the radiative return process.
% %
Section~\ref{SecI} presents the setup of the radiative return process together with a brief overview of the pion-photon interactions, with particular emphasis on the GVMD model. 
Section~\ref{subsec:penta} discusses the implementation strategy of the GVMD and the validation of the corrections to the radiative return process.
Section~\ref{SecII} investigates how the GVMD modifications affect different phenomenological settings, and highlights the nature of the results. The role of these effects is also studied on experimental data. 
Conclusions summarizing our findings and outlining directions for future work are provided in Section~\ref{sec:conclusions}.

% \section{GVMD in the radiative return}\label{SecI}
\section{The radiative return process}
\label{SecI}

\begin{figure}[t]
    %\centering
    \includegraphics[scale=0.7]{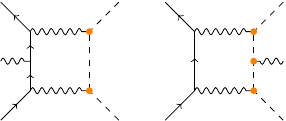}\quad\
    \includegraphics[scale=0.7]{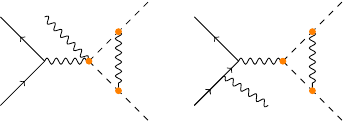}
     \caption{Representative diagrams for the two groups, TPE and FSC, changing under GVMD. The groups are further split into ISR and FSR contributions, corresponding from left to right to $\text{TPE}_\text{ISR}$, $\text{TPE}_\text{FSR}$, $\text{FSC}_\text{FSR}$ and $\text{FSC}_\text{ISR}$. }
     \label{fig:piondiags}
\end{figure}

We focus our attention on the radiative return process $e^+(p_1)e^-(p_2) \to \pi^+(p_3)\pi^-(p_5)\gamma(p_4)$, whose tree-level contribution takes the form, 
\begin{align}
    \mathcal{A}^{(0)} = A^{(0)}_{\text{ISR}}+ A^{(0)}_{\text{FSR}}\,,
\end{align}
where the subscripts account for the emission of the hard photon from the leptonic lines, initial state radiation (ISR), and from the hadronic lines, final state radiation (FSR). 
One can perform perturbative corrections on the electronic and hadronic contributions. By following the convention of Ref.~\cite{Aliberti:2024fpq}, we respectively refer to them as initial state corrections~(ISC) and final state corrections~(FSC). For instance, one-loop corrections to the radiative return process can be expressed as,
\begin{equation}
\mathcal{A}^{(1)} = \sum_{\mathcal{X} \in \{\text{ISR,FSR}\}}  A^{(1)}_{\text{ISC;}\mathcal{X}}+A^{(1)}_{\text{FSC};\mathcal{X}}
%\notag\\
+A^{(1)}_{\text{TPE;}\mathcal{X}}
%+A_{\text{VP;}\mathcal{X}}^\text{(1)}
\,,
\end{equation}
where the TPE term accounts for the two-photon-exchange diagrams. Note that additional vacuum-polarization contributions are included in the definition of the VFF.
%For every group of diagrams, the label ISR and FSR denotes the origin of the hard photon. 
In the standard \FxsQED approach, the pion VFF evaluated at
$s_{12}$ multiplies all FSR contributions as calculated with sQED, whereas all ISR contributions are weighted by $F_\pi(s_{35})$, with $s_{ij}~=~(p_i+p_j)^2$. It is useful to define:
\begin{align}
    \eta(\mathcal{X}) =
\begin{cases}
s_{35}, & \text{if } \mathcal{X}=\mathrm{ISR}\,,\\
s_{12}, & \text{if } \mathcal{X}=\mathrm{FSR}\,.
\end{cases}
\end{align} 
In GVMD, however, the VFF is inserted at the scalar--photon vertices, thereby modifying the standard sQED Feynman rules to:
%\tikzfeynmanset{compat=1.1.0}
%
% \begin{align}
% \begin{tikzpicture}[baseline={(current bounding box.center)},scale=0.75,transform shape]
% \begin{feynman}
%   \vertex (i) at (-1.2, 0) {};
%   \vertex[blob, minimum size=10pt,draw=orange] (a) at (0,0) {};
%   \vertex (o1) at (2.0,  0.8) {};
%   \vertex (o2) at (2.0, -0.8) {};
%   \diagram*{
%     (i)  -- [boson,  edge label={$q^\mu$}] (a),
%     (a)  -- [scalar, edge label={$p_1$}]  (o1),
%     (a)  -- [scalar, edge label'={$p_2$}] (o2),
%   };
% \end{feynman}
% \end{tikzpicture}
% = i\,Q_{\pi}(p_1-p_2)^\mu\,\textcolor{orange}{F_{\pi}}(q^2)\,,\label{eq:vertex1}\\[0.9em]
% \begin{tikzpicture}[baseline={(current bounding box.center)},scale=0.75,transform shape]
% \begin{feynman}
%   \vertex (i1) at (-1.2,  0.6) {};
%   \vertex (i2) at (-1.2, -0.6) {};
%   \vertex[blob, minimum size=10pt,draw=orange] (a) at (0,0) {};
%   \vertex (o1) at (2.0,  0.8) {};
%   \vertex (o2) at (2.0, -0.8) {};
%   \diagram*{
%     (i1) -- [boson,  edge label={$q_1^\mu$}]  (a),
%     (i2) -- [boson,  edge label'={$q_2^\nu$}] (a),
%     (a)  -- [scalar, edge label={$p_1$}]      (o1),
%     (a)  -- [scalar, edge label'={$p_2$}]     (o2),
%   };
% \end{feynman}
% \end{tikzpicture}
% = 2i\,Q_{\pi}^2g^{\mu\nu}\,\textcolor{orange}{F_{\pi}}(q_1^2)\textcolor{orange}{F_{\pi}}(q_2^2)\,.
% \label{eq:vertex2}
% \end{align}
%
\begin{align}
\parbox{18mm}{
\begin{tikzpicture}
\coordinate (e1) at (0.,0.);
\coordinate (e3) at (0.5,0);
\coordinate (e4) at (1,0.5);
\coordinate (e5) at (1,-0.5);
\draw[photon,  thin] (e1) -- (e3);
\draw[scalar,  thin] (e4) -- (e3);
\draw[scalar,  thin] (e5) -- (e3);
\node  at (-0.1,0.1) {\tiny$q^\mu$}; 
\node  at (1.2,0.5) {\tiny$p_1$}; 
\node  at (1.2,-0.5) {\tiny$p_2$}; 
\end{tikzpicture}
}
\to&\
\parbox{12mm}{
\begin{tikzpicture}
\coordinate (e1) at (0.,0.);
\coordinate (e3) at (0.5,0);
\coordinate (e4) at (1,0.5);
\coordinate (e5) at (1,-0.5);
\draw[photon,  thin] (e1) -- (e3);
\draw[scalar,  thin] (e4) -- (e3);
\draw[scalar,  thin] (e5) -- (e3);
\draw[draw=orange,fill=orange] (e3) circle (.05cm);
\end{tikzpicture}
}
=
i\,Q_{\pi}(p_1-p_2)^\mu\,\textcolor{orange}{F_{\pi}}(q^2)\,,\\
%\end{align*}
%
%
%\begin{align*}
\parbox{18mm}{
\begin{tikzpicture}
\coordinate (e1) at (0.,0.5);
\coordinate (e2) at (0.,-0.5);
\coordinate (e3) at (0.5,0);
\coordinate (e4) at (1,0.5);
\coordinate (e5) at (1,-0.5);
\draw[photon,  thin] (e1) -- (e3);
\draw[photon,  thin] (e2) -- (e3);
\draw[scalar,  thin] (e4) -- (e3);
\draw[scalar,  thin] (e5) -- (e3);
\node  at (-0.2,0.5) {\tiny$q_1^\mu$}; 
\node  at (-0.2,-0.5) {\tiny$q_2^\nu$}; 
\node  at (1.2,0.5) {\tiny$p_1$}; 
\node  at (1.2,-0.5) {\tiny$p_2$}; 
\end{tikzpicture}
}
\to&\
\parbox{12mm}{
\begin{tikzpicture}
\coordinate (e1) at (0.,0.5);
\coordinate (e2) at (0.,-0.5);
\coordinate (e3) at (0.5,0);
\coordinate (e4) at (1,0.5);
\coordinate (e5) at (1,-0.5);
\draw[photon,  thin] (e1) -- (e3);
\draw[photon,  thin] (e2) -- (e3);
\draw[scalar,  thin] (e4) -- (e3);
\draw[scalar,  thin] (e5) -- (e3);
\draw[draw=orange,fill=orange] (e3) circle (.05cm);
\end{tikzpicture}
}
= 2i\,Q_{\pi}^2g^{\mu\nu}\,\textcolor{orange}{F_{\pi}}(q_1^2)\textcolor{orange}{F_{\pi}}(q_2^2)\,.\notag
\end{align}
The vertices thus carry, on top of the expected Lorentz structure inherited in sQED, a VFF that depends on the momentum of the photons. If these carry a definite momentum, $F_\pi$ can be directly evaluated, yielding the usual treatment of \FxsQED. Nevertheless, for a set of diagrams, the VFF needs to be evaluated for loop momenta. These relevant one-loop diagrams, modified under the GVMD Feynman rules, correspond to the FSC and TPE groups. A representative diagram for each of them is displayed in Fig.~\ref{fig:piondiags}. We do not modify FSC corrections with GVMD. Overall, the corrections  are expected to be small~\cite{Aliberti:2024fpq, Budassi:2024whw,Fang:2025mhn}, and, at least for scan processes, are well approximated by \FxsQED\cite{Budassi:2024whw}. On the contrary, we do modify the TPE diagrams.

% \begin{figure*}[t]
%     \centering
%     \includegraphics[width=\textwidth]{Figures/Figure1.png}
%     \caption{Representative diagrams for the two groups, TPE and FSC, changing under GVMD. The groups are further split into ISR and FSR contributions, corresponding from left to right to $\text{TPE}_\text{ISR},\text{TPE}_\text{FSR},\text{FSC}_\text{ISR}$ and $\text{FSC}_\text{FSR}$. }
%     \label{fig:piondiags}
% \end{figure*}

%\subsection{The TPE diagrams}\label{subsec:penta}
\section{The TPE diagrams within GVMD}\label{subsec:penta}

The TPE contributions can be broken down into $A^{(1)}_\text{TPE;ISR}$ and $A^{(1)}_\text{TPE;FSR}$ (see Fig.~\ref{fig:piondiags}). 
We calculate both under the GVMD modifications. 
%As already hinted in the Introduction, this might not be optimal in the second group of charges 
For the first set of diagrams $A^{(1)}_\text{TPE;ISR}$, GVMD is a good approximation to the full Compton tensor, $\gamma^*\gamma^*\to\pi^+\pi^-$, and is an improvement over the usual \FxsQED description. We will also refer to it as GVMD$_\text{ISR}$, and always show it by itself in the results of Section~\ref{SecII}. It is unclear if GVMD is a valid approximation to describe the other set of diagrams $A^{(1)}_\text{TPE;FSR}$. As recently shown in the photon soft limit approximation~\cite{citation-key}, \FxsQED is a valid first approximation to the $\gamma^*\gamma^*\gamma\to\pi^+\pi^-$ process. Thus, it is plausible that GVMD is also an improvement in this case. Consequently, both as a natural extension to GVMD$_\text{ISR}$, and to prepare for the future, we decide to also include $A^{(1)}_\text{TPE;FSR}$, with the option of turning it on and off left to the user of the code (see Section~\ref{SecII} and Appendix~\ref{Appendix} for additional details).

We perform the calculation of the amplitudes working in dimensional regularization. Following~\cite{Ignatov:2022iou}, we parametrize the VFF as a sum of Breit--Wigner terms, 
\begin{align}\label{eq:FF}
    F_\pi(q^2) = \sum_{v} a_v\left(1+\frac{q^2}{\Lambda_v-q^2}\right)\,,
\end{align}
with the constraint that $F_\pi(0)=1$, and thus the constants $a_v$ need to satisfy $\sum_v a_v =1$. The complex parameters $\Lambda_v = m_v^2-im_v\Gamma_v$ are fixed by the mass $m_v$ and width $\Gamma_v$ of the vector meson $v$. The complex parameters $\Lambda_v$ can be treated as effective masses, which will be acquired by the virtual photon of the interaction. In this way, once both VFF are expanded, we can separate the whole amplitude in three individual contributions, namely contributions that contain only massless photons, $\delta_\text{NN}$, equivalent to the sQED contribution, those that both virtual photons are effectively massive, $\delta_\text{MM}(\Lambda_v,\Lambda_w)$, and finally the contributions with only one massive photon, $\delta_\text{MN}(\Lambda_v)$. This decomposition can be depicted diagrammatically, in the $\mathcal{X}=\text{ISR}$ case as, 
\begin{widetext}
\begin{align}
\label{eq:gvmd_to_sqed}
\parbox{24mm}{
\begin{tikzpicture}	
\coordinate (e1) at (0.,0.5);
\coordinate (e2) at (0.,2);
\coordinate (v1) at (0.5,0.75);
\coordinate (v2) at (0.5,1.75);
\coordinate (v3) at (1.2,0.5);
\coordinate (v4) at (1.2,2);
\coordinate (v7) at (0.5,1.225);
\draw[photon,  thin] (e1) -- (v1);
\draw[photon,  thin] (e2) -- (v2);
\draw[scalar,  thin] (v1) -- (v3);
\draw[scalar,  thin] (v2) -- (v4);
\fill[gray](v7) ellipse (1.5ex and 4ex) ;
\node[white] at (v7) {\tiny \Longstack{G V M D}};
\node  at (-0.2,0.6) {\small$\gamma^*$}; 
\node  at (-0.2,2.1) {\small$\gamma^*$}; 
\node  at (1.5,0.6) {\small$\pi^+$}; 
\node  at (1.5,2.1) {\small$\pi^-$}; 
\end{tikzpicture}
}
=\
\underbrace{
\parbox{15mm}{\
\begin{tikzpicture}
\coordinate (e1) at (0.,0.5);
\coordinate (e2) at (0.,2);
\coordinate (v1) at (0.5,0.75);
\coordinate (v2) at (0.5,1.75);
\coordinate (v3) at (1.2,0.5);
\coordinate (v4) at (1.2,2);
\coordinate (v7) at (0.5,1.225);
\draw[photon, thin] (e1) -- (v1);
\draw[photon, thin] (e2) -- (v2);
\draw[scalar,  thin] (v1) -- (v3);
\draw[scalar,  thin] (v2) -- (v4);
\fill[gray](v7) ellipse (1.5ex and 4ex) ;
\node[white] at (v7) {\tiny \Longstack{s Q E D}};
\end{tikzpicture}
}
}_{\delta_\text{NN}}
-\sum_{v}a_v\underbrace{\left(\!\!
\parbox{20mm}{
\begin{tikzpicture}
\coordinate (e1) at (0.,0.5);
\coordinate (e2) at (0.,2);
\coordinate (v1) at (0.5,0.75);
\coordinate (v2) at (0.5,1.75);
\coordinate (v3) at (1.2,0.5);
\coordinate (v4) at (1.2,2);
\coordinate (v7) at (0.5,1.225);
\draw[photon,  very thick] (e1) -- (v1);
\draw[photon,  thin] (e2) -- (v2);
\draw[scalar,  thin] (v1) -- (v3);
\draw[scalar,  thin] (v2) -- (v4);
\fill[gray](v7) ellipse (1.5ex and 4ex);
\node  at (-0.25,0.5) {\small$\Lambda_v$}; 
\node[white] at (v7) {\tiny \Longstack{s Q E D}};
\end{tikzpicture}
}
+\
\parbox{20mm}{
\begin{tikzpicture}
\coordinate (e1) at (0.,0.5);
\coordinate (e2) at (0.,2);
\coordinate (v1) at (0.5,0.75);
\coordinate (v2) at (0.5,1.75);
\coordinate (v3) at (1.2,0.5);
\coordinate (v4) at (1.2,2);
\coordinate (v7) at (0.5,1.225);
\draw[photon, thin] (e1) -- (v1);
\draw[photon, very thick] (e2) -- (v2);
\draw[scalar,  thin] (v1) -- (v3);
\draw[scalar,  thin] (v2) -- (v4);
\fill[gray](v7) ellipse (1.5ex and 4ex) ;
\node[white] at (v7) {\tiny \Longstack{s Q E D}};
\node  at (-0.25,2) {\small$\Lambda_v$}; 
\end{tikzpicture}
}
\!\right)
}_{\delta_\text{MN}(\Lambda_v)}
+
\sum_{v,w}a_va_w\
\underbrace{
\parbox{20mm}{
\begin{tikzpicture}
\coordinate (e1) at (0.,0.5);
\coordinate (e2) at (0.,2);
\coordinate (v1) at (0.5,0.75);
\coordinate (v2) at (0.5,1.75);
\coordinate (v3) at (1.2,0.5);
\coordinate (v4) at (1.2,2);
\coordinate (v7) at (0.5,1.225);
\draw[photon, very thick] (e1) -- (v1);
\draw[photon, very thick] (e2) -- (v2);
\draw[scalar,  thin] (v1) -- (v3);
\draw[scalar,  thin] (v2) -- (v4);
\fill[gray](v7) ellipse (1.5ex and 4ex) ;
\node[white] at (v7) {\tiny \Longstack{s Q E D}};
\node  at (-0.25,0.5) {\small$\Lambda_w$}; 
\node  at (-0.25,2) {\small$\Lambda_v$}; 
\end{tikzpicture}
}
}_{\delta_\text{MM}(\Lambda_v,\Lambda_w)}
\,.
\end{align}
\end{widetext}
The GVMD amplitude of the TPE diagrams is thus easy to write down as,
\begin{align}
A^{(1)}_{\text{TPE;}\mathcal{X}} &=  \phantom{+}\delta_\text{NN}^\mathcal{X} 
\\
&+\sum_{v} a_v \left[\delta_\text{MN}^\mathcal{X}(\Lambda_v)
 \notag
+\sum_w a_w \delta_\text{MM}^\mathcal{X}(\Lambda_v,\Lambda_w)\right]\,.
\end{align}
This works for both $\mathcal{X}\in\{$ISR, FSR\}, and allows one to calculate any general VFF Breit--Wigner fit. It is of interest to note that the GVMD modifications do not introduce ultraviolet (UV) singularities and leave unchanged the infrared (IR) structure of the one-loop contribution predicted by lower \FxsQED amplitudes in these diagrams~\cite{Catani:2000ef}.
Specifically, $\delta_\text{NN},\delta_\text{MN},\delta_\text{MM}$ are individually UV finite, and the 
IR divergent singularities come only from $\delta_\text{NN},\delta_\text{MN}$, with $\delta_\text{MM}$ a fully finite contribution. Hence, we can split the GVMD contribution as\footnote{
There is a slight abuse in the notation adopted here. Within the GVMD framework, the quantity actually computed corresponds to the interference between the relevant subset of TPE diagrams and the Born amplitude, which introduces factors of $F_\pi^\dagger(\eta(\mathcal{X}))$ from the Born amplitude.}, 
% \Pau{Should we unify the notation with eq.2?}
% \WJT{Here we can mention the interference, or what do you have in mind?}\Pau{I am not quite sure. The tree level will carry with itself a FF at different energies depending on if they are an ISR tree level or FSR. I don't think it is clear including this here. Let me know your opinion on the footnote I added}
\begin{equation}
\mathcal{A}_{\text{TPE;}\mathcal{X}}^\text{GVMD}
= \delta_\text{\FxsQED}^\mathcal{X} + \delta_\text{FF}^\mathcal{X}\,,
\end{equation}
with
\begin{equation}
\begin{aligned} \label{eq:amp}
\delta_\text{\FxsQED}^\mathcal{X} &= F_\pi(\eta(\mathcal{X}))\,\delta_\text{NN}^\mathcal{X}\,,\\
\delta_\text{FF}^\mathcal{X} &=
\sum_{v} a_v \Bigg[
\delta_\text{MN}^\mathcal{X}(\Lambda_v)
-\frac{\eta(\mathcal{X})}{\Lambda_v-\eta(\mathcal{X})}\,\delta_\text{NN}^\mathcal{X} \\
&\hspace{2.2em}
+\sum_w a_w\,\delta_\text{MM}^\mathcal{X}(\Lambda_v,\Lambda_w)
\Bigg]\,.
\end{aligned}
\end{equation}
If $\delta_\text{FF}=0$, one easily recovers \FxsQED. In order to avoid complications related to different regularization schemes in different MC generators, and make 
our calculation easy to interface, we provide \texttt{Fortran} routines to evaluate $\delta_\text{FF}$ in the ancillary files~\cite{zenodo}, and give instructions on how to use them in Appendix~\ref{Appendix}. Note that $\delta_\text{FF}$ is defined such that it is IR finite, concentrating the IR structure in $\delta_\text{\FxsQED}$, which automatically matches the real corrections.

The calculation of the amplitudes of $\delta_\text{FF}$ is performed with two different methods, to compare and ensure the correctness of the result. In one, a dedicated pipeline with the combination of \texttt{qgraf}~\cite{Nogueira:1991ex}, \texttt{Tapir}~\cite{Gerlach:2022qnc}, and \texttt{Form}~\cite{Vermaseren:2000nd,Ruijl:2017dtg,Kuipers:2012rf,Davies:2026cci} is used to generate, dress and evaluate the topologies, directly with the GVMD Feynman rules, which are implemented in a custom model in \texttt{Tapir}. 
In the other, the needed amplitudes are calculated with \textsc{FeynArts}~\cite{Hahn:2000kx} and \textsc{FeynCalc}~\cite{Shtabovenko:2025lxq}. In this case, instead of modifying the sQED Feynman rules, the groups $\delta_\text{MM}$ and $\delta_\text{MN}$ are obtained from $\delta_\text{NN}$, by means of a ``massification'' of the loop propagators, $\mathcal{A}_\text{TPE}^\text{sQED}\to \mathcal{A}_\text{TPE}^\text{GVMD}$, closely following~\eqref{eq:gvmd_to_sqed}. Agreement is found between methods.

In addition, we test two complementary strategies for evaluating the amplitudes.
The first strategy employs integration-by-parts (IBP) identities~\cite{Tkachov:1981wb,Chetyrkin:1981qh}, generated with
\texttt{Kira}~\cite{Klappert:2020nbg}, to reduce all loop integrals to a set of
master integrals~\cite{Laporta:2000dsw}. Following the basis suggested by the rotation matrix of the
differential equation constructed in~\cite{PetitRosas:2025xhm}, we isolate the
IR singularity and thereby evaluate the remaining Feynman integrals
directly in strictly four space-time dimensions, with \texttt{Collier}~\cite{Denner:2002ii,Denner:2016kdg}.
The second strategy follows the conventional tensor-integral approach: the
one-loop tensor integrals (up to rank~2) are reduced to the standard
Passarino--Veltman basis and are likewise evaluated with \texttt{Collier}.

Although the IBP reduction substantially decreases the number of integrals, the
increased algebraic complexity of the kinematic prefactors generated by the IBP
identities leads to an overall worse performance, by roughly a factor of~5,
relative to the second strategy. In the latter, the $\mathcal{O}(300)$ scalar
integrals dominate the CPU time, while the evaluation of the kinematic
coefficients is negligible, especially after further optimization with
\texttt{Form}.

Finally, we find that organizing the amplitudes into the
$\delta_\text{NN}$, $\delta_\text{MN}$, and $\delta_\text{MM}$ components
reduces CPU time by enabling extensive reuse of \texttt{Collier}'s caching:
the same loop functions are evaluated repeatedly for different $\Lambda$
values, rather than recomputing the full amplitude for each choice.
\\   

\section{Results}
\label{SecII}

We study the impact of the GVMD modifications for a set of scenarios. To do so, we interface the GVMD code to the MC generator \texttt{Phokhara}. With it, we explore and discuss all benchmark scenarios presented in~\cite{Aliberti:2024fpq}. Then, we will review a relaxed KLOE large angle  scenario used to study the forward--backward asymmetry, following the investigations performed in~\cite{WorkingGrouponRadiativeCorrections:2010bjp}. In order to describe KLOE data in the forward--backward asymmetry scenario, additional contributions are needed. These are described in Subsection~\ref{subsec:fba}, and more in detail in Appendix~\ref{Appendix:ISC}. Finally, to eliminate biases from using different parametrizations of $F_\pi$, we stick to using the one defined in~\cite{Ignatov:2022iou} for all scenarios.
%, and turn off vacuum polarizations. 

To apply cuts to the scenarios and study a range of differential cross section distributions, we define the kinematic variables $\theta_\gamma$, $\theta_\pm$, which are the photon and pion angles, the hard photon energy, $E_\gamma$, and the pion perpendicular and azimuthal momenta, $p_z^\pm$, $p_\perp^\pm$, as well as their invariant mass $M_{\pi\pi}$. They are all defined in accordance with the consensus established in~\cite{Aliberti:2024fpq}. Furthermore, we give the forward--backward asymmetry a definition such that
\begin{equation}\label{eq:fbasymmetry}
    \mathcal{A}_\text{FB}(X,Y)=\frac{\frac{d\sigma }{dX}(Y>90^\circ)-\frac{d\sigma }{dX}(Y<90^\circ)}{\frac{d\sigma }{dX}(Y>90^\circ)+\frac{d\sigma }{dX}(Y<90^\circ)}\,,
\end{equation}
where $X$ account for kinematic variables, and $Y$ must strictly be an angular one.

\subsection{KLOE-like large angle}\label{subsec:kloela}

\begin{figure*}[t]
    \centering
    \includegraphics[width=0.5\textwidth]{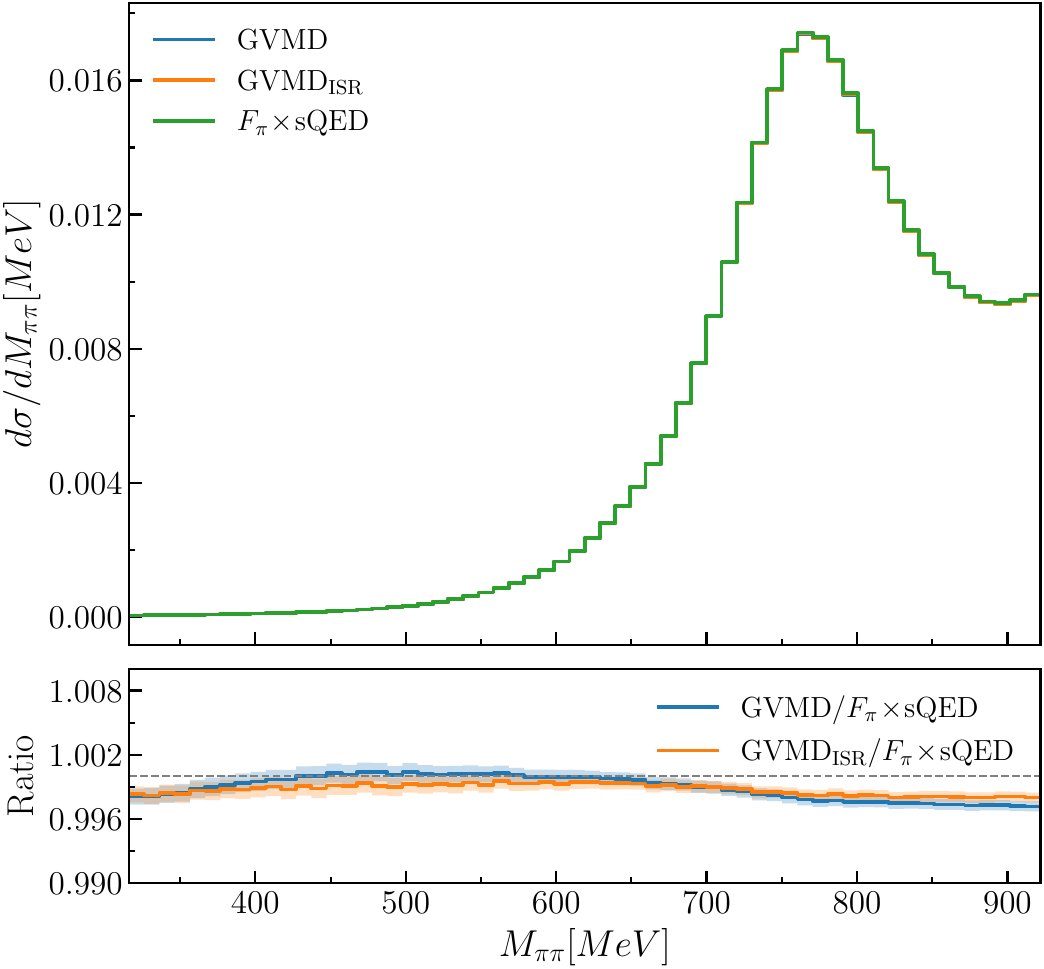}\hfill
    \includegraphics[width=0.5\textwidth]{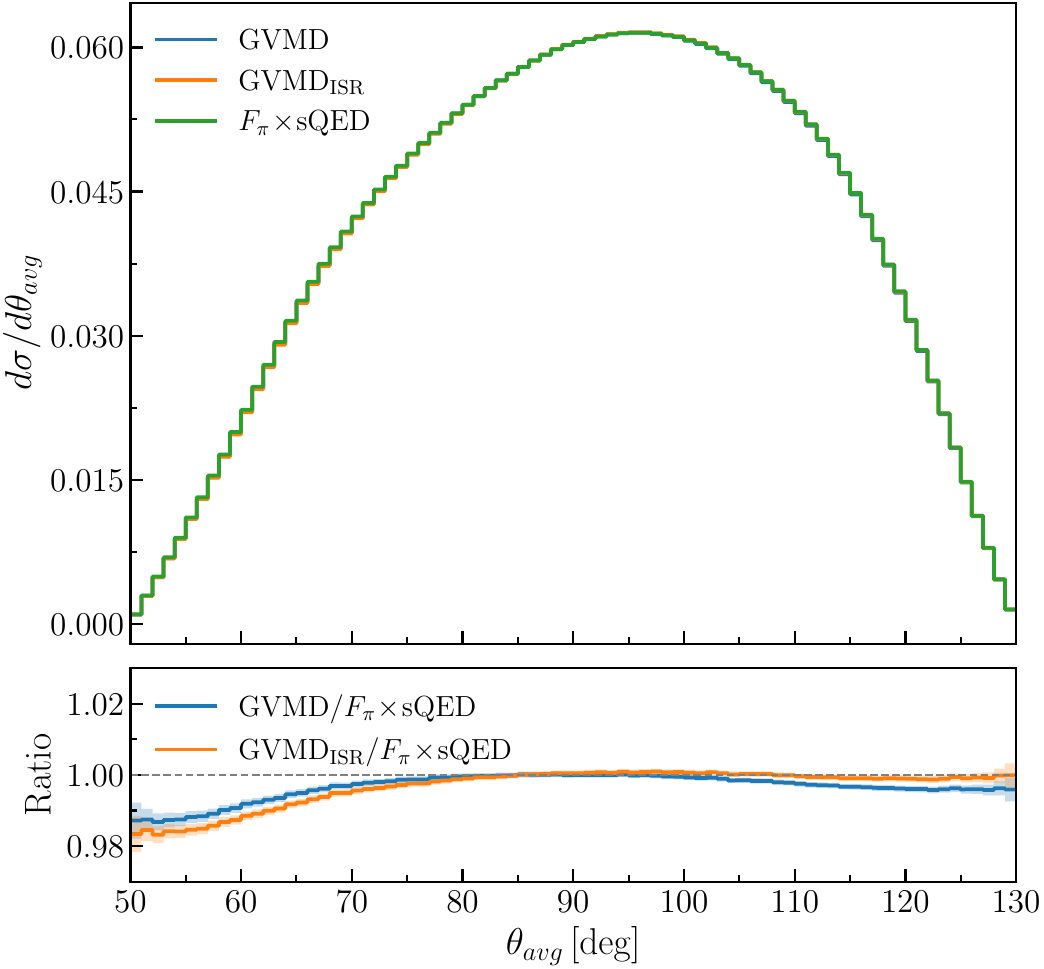}
    \caption{Differential cross section versus pion--pair invariant mass (left) and $\theta_{\text{avg}}$ (right) under \FxsQED, the full GVMD modifications and only the ISR diagrams modified with GVMD. The bottom panel of each plot shows, the ratio between the modifications and \FxsQED. Colour bands correspond to statistical uncertainties. Note the difference between y-axis scales. }
    \label{fig:kloeI}
\end{figure*}

We test the GVMD modifications in a KLOE-like large angle (LA) scenario. Specifically, we apply the following cuts:
\begin{equation}
\begin{aligned}
& \sqrt{s}=1.02~\text{GeV}\,,~50^\circ \le \theta_{\pm} \le 130^\circ\,,\\[3pt] 
& E_{\gamma} > 20~\mathrm{MeV}\,,~~~
50^\circ \le \theta_{\gamma} \le 130^\circ\,,\\[3pt]
& |p_{z}^{\pm}| > 90~\mathrm{MeV}\; \lor\; p_{\perp}^{\pm} > 160~\mathrm{MeV}\,,\\[3pt]
&0.1~\mathrm{GeV}^2 \le M_{\pi\pi}^{2} \le 0.85~\mathrm{GeV}^2\,.\\
\end{aligned}
\end{equation}

We present the comparison between GVMD and \FxsQED in Fig.~\ref{fig:kloeI}, where we display the differential cross section for the scenario, as plotted versus the invariant mass of the pions, $M_{\pi\pi}$, and $\theta_\text{avg}=(\theta_-+\theta_++\pi)/2$. We split the GVMD contribution in the total one and GVMD$_{\text{ISR}}$. Sub-percent effects are seen in the invariant mass distributions, with maximum of $\sim0.3\%$, with no particular trend, and maximal deviation in the edges of the distribution. A larger, at the percent level, divergence is observed in the angular distributions. Even in the full GVMD, which tames the pure GVMD$_\text{ISR}$, the corrections reach more than one percent in $\theta_\text{avg}=50^\circ$. As the corrections are asymmetric over the angular distributions, the cross-section, seen in Table~\ref{table:cs}, also differs between approaches, with a difference of roughly $0.2\%$. Due to the biggest contribution to $a_\mu^\text{HVP}$ being the $\rho$ peak region, we also show in Table~\ref{table:csrho} the cross section of the region  0.5 GeV$^2 < M_{\pi\pi}^2 < 0.7$ GeV$^2$. %The GVMD modifications also have a direct impact on the forward--backward asymmetry. To illustrate it, we plot in the left panel of Fig.~\ref{fig:asymetry} the distribution $\mathcal{A}_{FB}(\theta_\text{avg},\theta_\text{avg})$, with the difference between approaches shown in the bottom panel. The GVMD modifications yield a charge asymmetry $0.005$ bigger in the point of maximal absolute difference.

%\begin{figure*}[t]
%    \centering
    %\includegraphics[width=0.5\textwidth]{Figures/AFBKloeI.pdf}\hfill
    %\includegraphics[width=0.5\textwidth]{Figures/AFBKloeII.pdf}
    %\caption{Forward--backward asymmetry for a KLOE-like large angle (left) and small angle (right) analysis, shown versus the angular variable $\theta_\text{avg}$, when evaluated with \texttt{Phokhara} using \FxsQED, a full GVMD calculation or only its ISR modifications. In the bottom panels, the difference between apporaches, defined such that $\Delta_\text{X} = \text{X} - $\FxsQED. }
    %\label{fig:asymetry}
%\end{figure*}

\begin{table}[t]
\centering
\begin{tabular}{lccc}
\hline
 & $\sigma_{F_\pi \times \mathrm{sQED}}$ & $\sigma_{\mathrm{GVMDISR}}$ & $\sigma_{\mathrm{GVMD}}$ \\
\hline
KLOE LA  & 3.2904(2)   & 3.2851(2)   & 3.2839(2)   \\
KLOE SA  & 16.3046(9)  & 16.3050(9)  & 16.3050(9)  \\
BESIII   & 0.08461(2)  & 0.08472(2)  & 0.08473(2)  \\
B        & 0.010173(4) & 0.010175(4) & 0.010176(4) \\
\hline
\end{tabular}
\caption{Cross sections for each scenario as modelled with \FxsQED, GVMD or only the GVMD modifications in ISR contributions. LA and SA denote, respectively, large and small angle scenarios.}
\label{table:cs}
\end{table}

\begin{table}[t]
\centering
\begin{tabular}{lccc}
\hline
 & $\sigma^\rho_{F_\pi \times \mathrm{sQED}}$ & $\sigma^\rho_{\mathrm{GVMDISR}}$ & $\sigma^\rho_{\mathrm{GVMD}}$ \\
\hline
KLOE LA  & 1.9038(2)   & 1.9010(2)   & 1.8997(2)   \\
KLOE SA  & 8.8450(8)  & 8.8453(8)  & 8.8452(8)  \\
BESIII   & 0.03908(1)  & 0.03915(1)  & 0.03915(1)  \\
B        & 0.004747(3) & 0.004748(4) & 0.004749(3) \\
\hline
\end{tabular}
\caption{Cross sections for each scenario in the $\rho$ peak region 0.5 GeV$^2 < M_{\pi\pi}^2 < 0.7$ GeV$^2$  as modelled with \FxsQED, GVMD or GVMD$_\text{ISR}$. LA and SA denote large and small angle scenarios.}
\label{table:csrho}
\end{table}

\subsection{KLOE-like small angle}
\label{subsec:kloesa}

\begin{figure*}[t]
    \centering
    \includegraphics[width=0.5\textwidth]{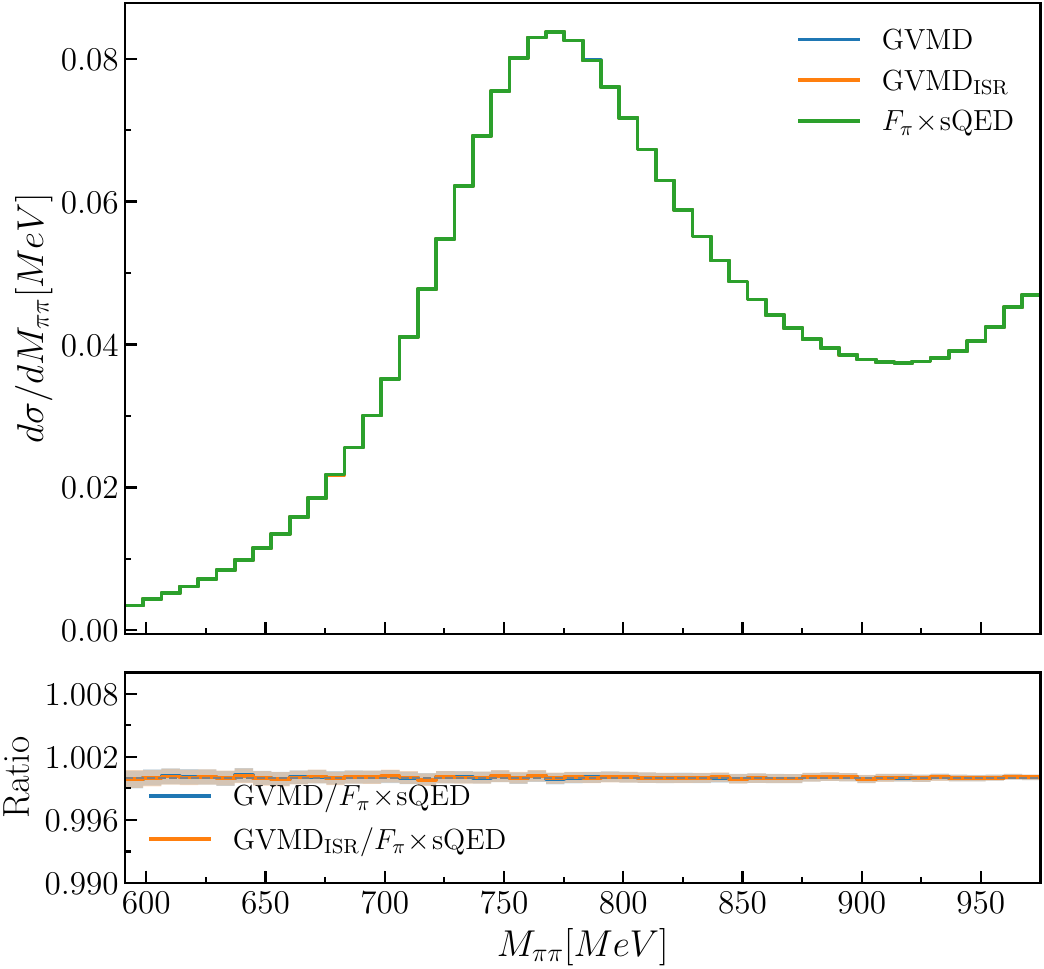}\hfill
    \includegraphics[width=0.5\textwidth]{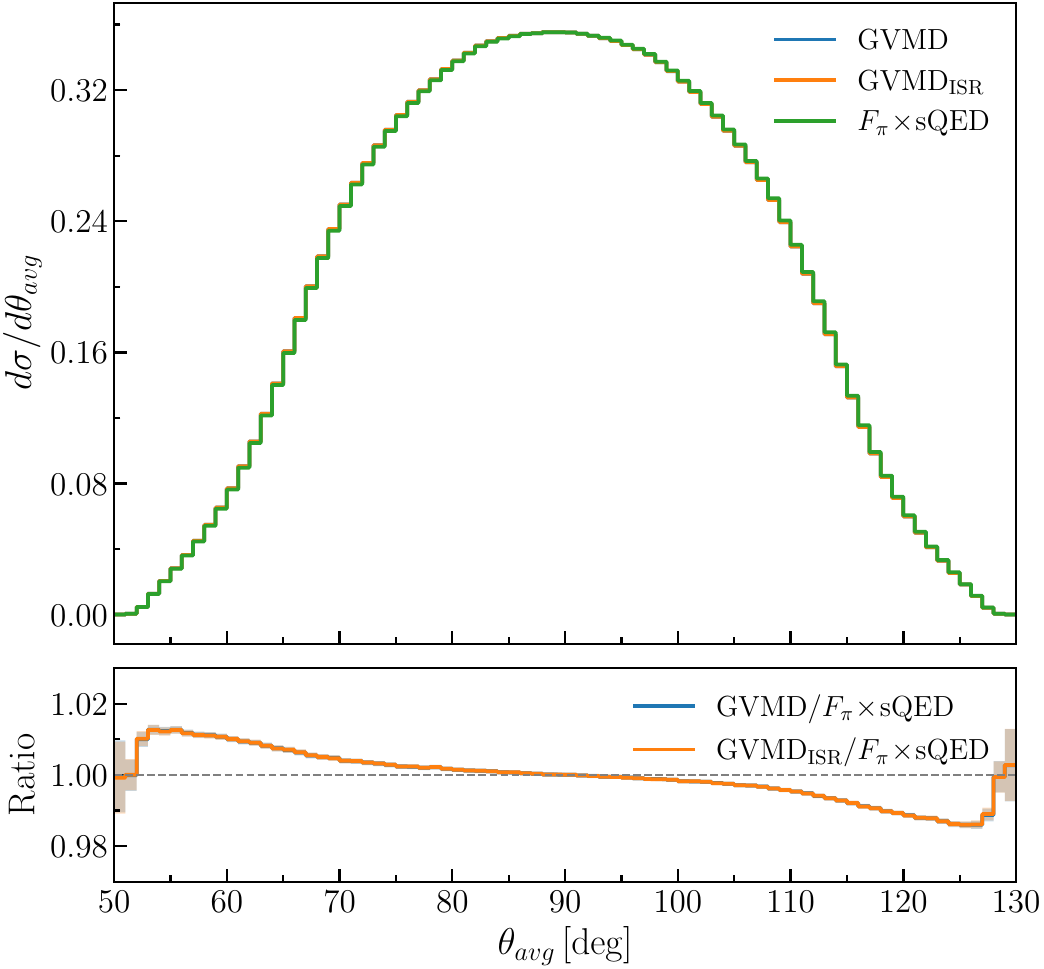}
    \caption{Differential cross section versus pion--pair invariant mass (left) and $\theta_{\text{avg}}$ (right) under \FxsQED, the full GVMD modifications and only the ISR diagrams modified with GVMD, under the KLOE-like small angle scenario. The ratio between the GVMD modifications and \FxsQED can be seen in the bottom of the plot.}
    \label{fig:kloeII}
\end{figure*}

We repeat the analysis for a KLOE-like small angle (SA) scenario. We now apply cuts:
\begin{equation}
\begin{aligned}
& \sqrt{s}=1.02~\text{GeV}\,,~50^\circ \le \theta_{\pm} \le 130^\circ\,,\\[3pt] 
& |p_{z}^{\pm}| > 90~\mathrm{MeV} \;\; \lor \;\; p_{\perp}^{\pm} > 160~\mathrm{MeV}\,,\\[3pt]
& \theta_{\widetilde{\gamma}} \le 15^\circ~~~\lor~~~
\theta_{\widetilde{\gamma}} > 165^\circ\,,\\[3pt]
&0.35~\mathrm{GeV}^2 \le M_{\pi\pi}^{2} \le 0.95~\mathrm{GeV}^2\,,\\
\end{aligned}
\end{equation}
where $\widetilde{\gamma}$ represents the polar angle of $\vec{p_{\widetilde{\gamma}}}~=~-(\vec{p_+}~+~\vec{p_-})$. The results for the differential cross section distributions are presented in Fig.~\ref{fig:kloeII}. In this case, GVMD has no observable effect on the invariant mass distribution, nor on any other distribution of a charge-even variable. The angular distributions, however, show a disagreement of roughly the same maximal size as in the large angle scenario, $2\%$, but with a trend that is symmetric over the phase space. The fact that no deviation is seen in C-even variables, and no impact is observed in the cross section values (also reported in Tables~\ref{table:cs} and~\ref{table:csrho}), points towards the main Feynman diagrams contributing to the angular divergence being C-odd. In particular, since the KLOE small angle scenario is dominated by ISR~\cite{Aliberti:2024fpq}, we conjecture that the modifications on the TPE$_\text{ISR}$, interfered with $A_\text{ISR}^{(0)}$, contribute the most. This is supported by the fact that no change is observed between the full GVMD corrections and only the ISR part.  

 %Even if the final cross section values show little to no effect, the impact on the experimental analysis must be carefully assessed. In particular, studies performed to evaluate systematic uncertainties arising from angular distributions should be cautiously revisited, as well as acceptance rates. 

\subsection{BESIII-like and B scenarios}
We present here the two remaining scenarios used in the radiative return studies of~\cite{Aliberti:2024fpq}. Even if they are different scenarios, we group them in the same Subsection for a simple reason. That is, the centre of mass energies of the colliders modelled in these two cases are significantly larger than the energy where $F_\pi$ peaks. We anticipate that this will greatly minimize the impact of the GVMD modifications, since the pion vector form factor rapidly falls from its peak. 

\begin{figure*}[t]
    \centering
    \includegraphics[width=0.5\textwidth]{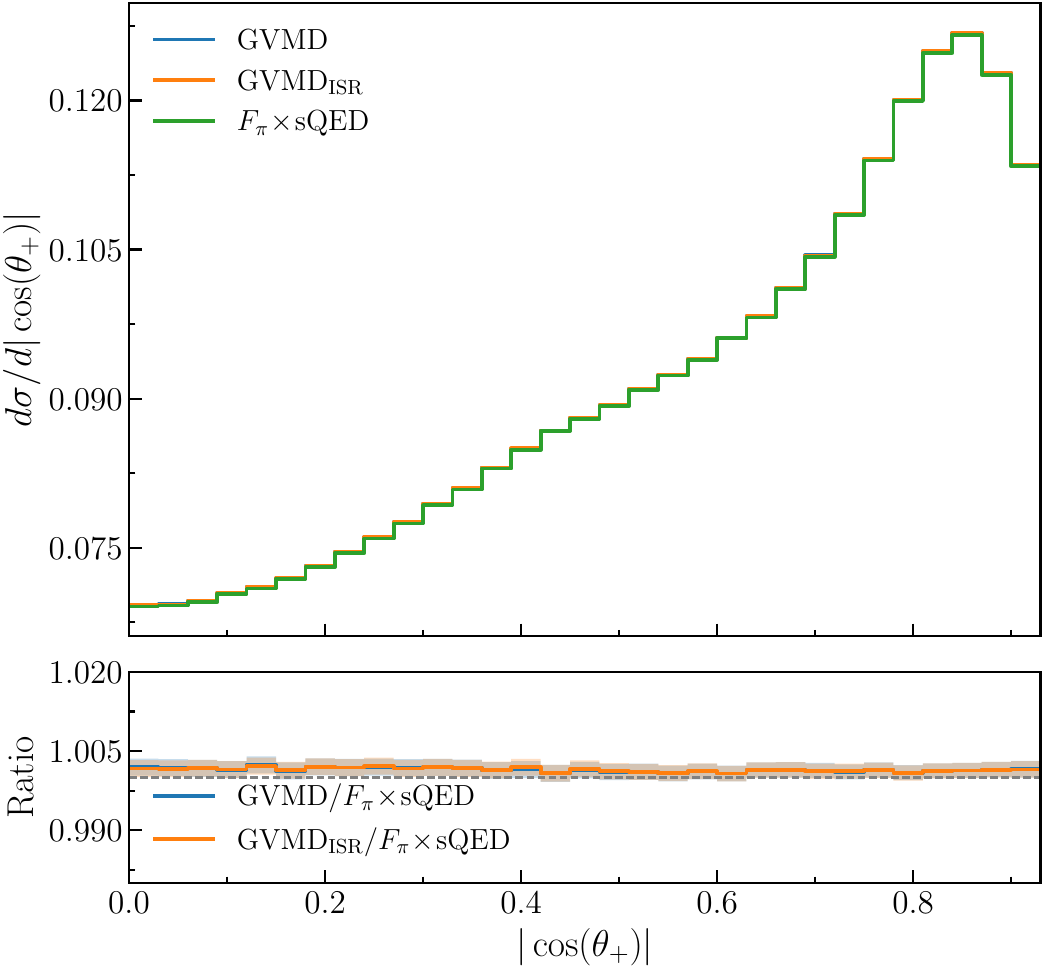}\hfill
    \includegraphics[width=0.5\textwidth]{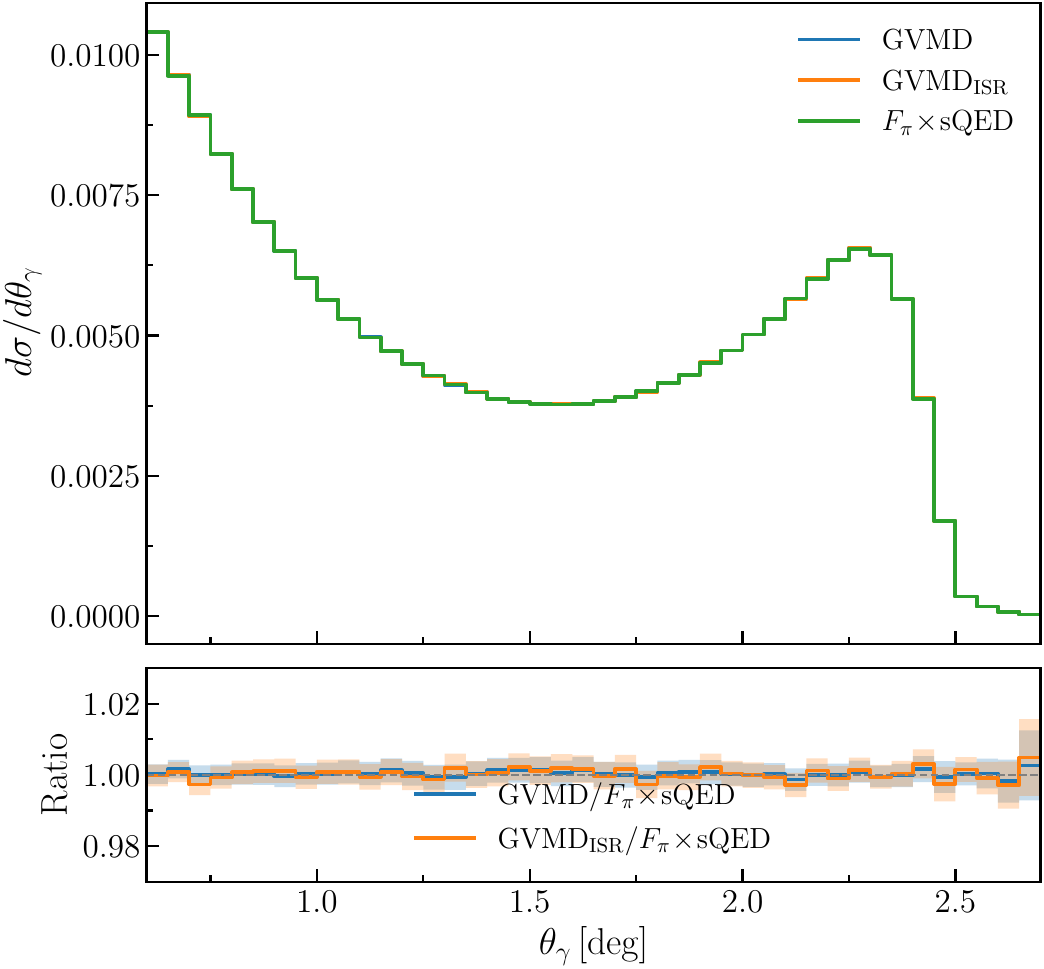}
    \caption{BESIII scenario $|\cos(\theta_+)|$ (left) and B scenario $\theta_\gamma$ (right) differential cross section with the full GVMD modifications, \FxsQED and only the ISR diagrams modified with GVMD. The ratio between the approaches is displayed in the bottom panel.}
    \label{fig:Bs}
\end{figure*}

The BESIII experiment runs at $\sqrt{s}=4$ GeV, and we apply the following cuts for the scenario:
\begin{equation}
\begin{aligned}
& |\cos\theta_\pm|<0.93\,\land\,p_\pm^\perp > 300~\mathrm{MeV}\,,\\[3pt] 
& ((|\cos\theta_\gamma|<0.8\land E_\gamma > 25~\mathrm{MeV})~\lor\\[3pt]
& (0.86 < |\cos\theta_\gamma|<0.92\land E_\gamma > 50~\mathrm{MeV}))\,,\\[3pt]
&\exists!\,\gamma~\text{with}\; E_\gamma \ge 400~\mathrm{MeV}\,.\\
\end{aligned}
\end{equation}
Alternatively, with a centre-of-mass energy roughly an order of magnitude larger than that of KLOE, the B scenario sets $\sqrt{s}=10$ GeV, with experimental cuts for the scenario being:
\begin{equation}
\begin{aligned}
& 0.65~\mathrm{rad} \le \theta_{\pm} \le 2.75~\mathrm{rad}
\;\wedge\;
p_{\pm} \ge 1~\mathrm{GeV}\,,\\[3pt] 
& 0.6~\mathrm{rad} \le \theta_{\gamma} \le 2.7~\mathrm{rad}
\;\wedge\;
E_{\gamma} \ge 3~\mathrm{GeV}\,,\\[3pt]
& \theta_{\tilde{\gamma},\gamma^{(h)}} \le 0.3~\mathrm{rad}\,,~~M_{XX\gamma} \ge 8~\mathrm{GeV}\,.\\
\end{aligned}
\end{equation}
We display the differential cross section for each scenario in Fig.~\ref{fig:Bs}, as seen versus angular variables. We note that a slight deviation of $0.2\%$ is seen in plots of both angular and charge even variables in the BESIII scenario. This impacts the cross section at the per-mille level (see Table~\ref{table:cs} and Table~\ref{table:csrho}). On the contrary, at $\sqrt{s}=10\text{ GeV}$ and with cuts from the B scenario, no difference between GVMD and \FxsQED is seen in any differential cross section distribution nor in the cross section, within the statistical error of the MC.   

\subsection{Forward--backward asymmetry}\label{subsec:fba}
In order to compare the forward--backward asymmetry in KLOE, we use the data from~\cite{WorkingGrouponRadiativeCorrections:2010bjp}.\footnote{In particular, we use the preliminary KLOE data presented in Figs. 53 and~54 of Ref~\cite{WorkingGrouponRadiativeCorrections:2010bjp}.} We run the same scenario, by applying a broad cut on the angle of the charged particles and hard photon, as well as a tight cut on the track mass, such that
\begin{equation}
\begin{aligned}
& 50^\circ \le \theta_{\gamma,\pm} \le 130^\circ\,,\\[3pt]
& | M_{\text{trk}} - M_\pm| < 10~\text{MeV}\,.\\
\end{aligned}
\end{equation}
Note that in this case we turn on additional ISC$_\text{FSR}$ contributions arising from the $\phi$ decays, which we refer to as ISC$_\phi$, and discuss in detail in Appendix~\ref{Appendix:ISC}. We will always plot them separately with either \FxsQED or GVMD.

\begin{figure*}[t]
    \centering
    \includegraphics[width=0.5\textwidth]{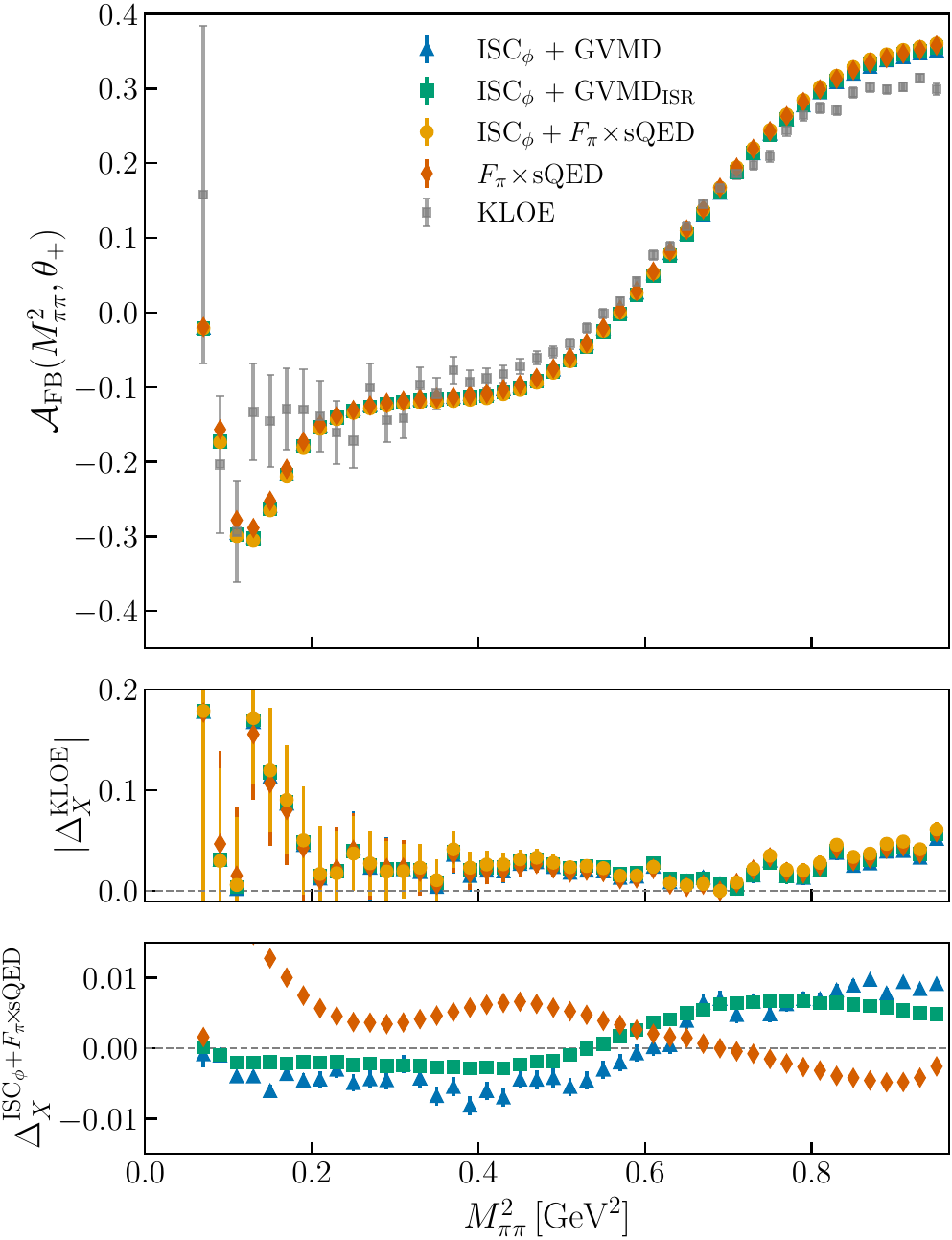}\hfill
    \includegraphics[width=0.5\textwidth]{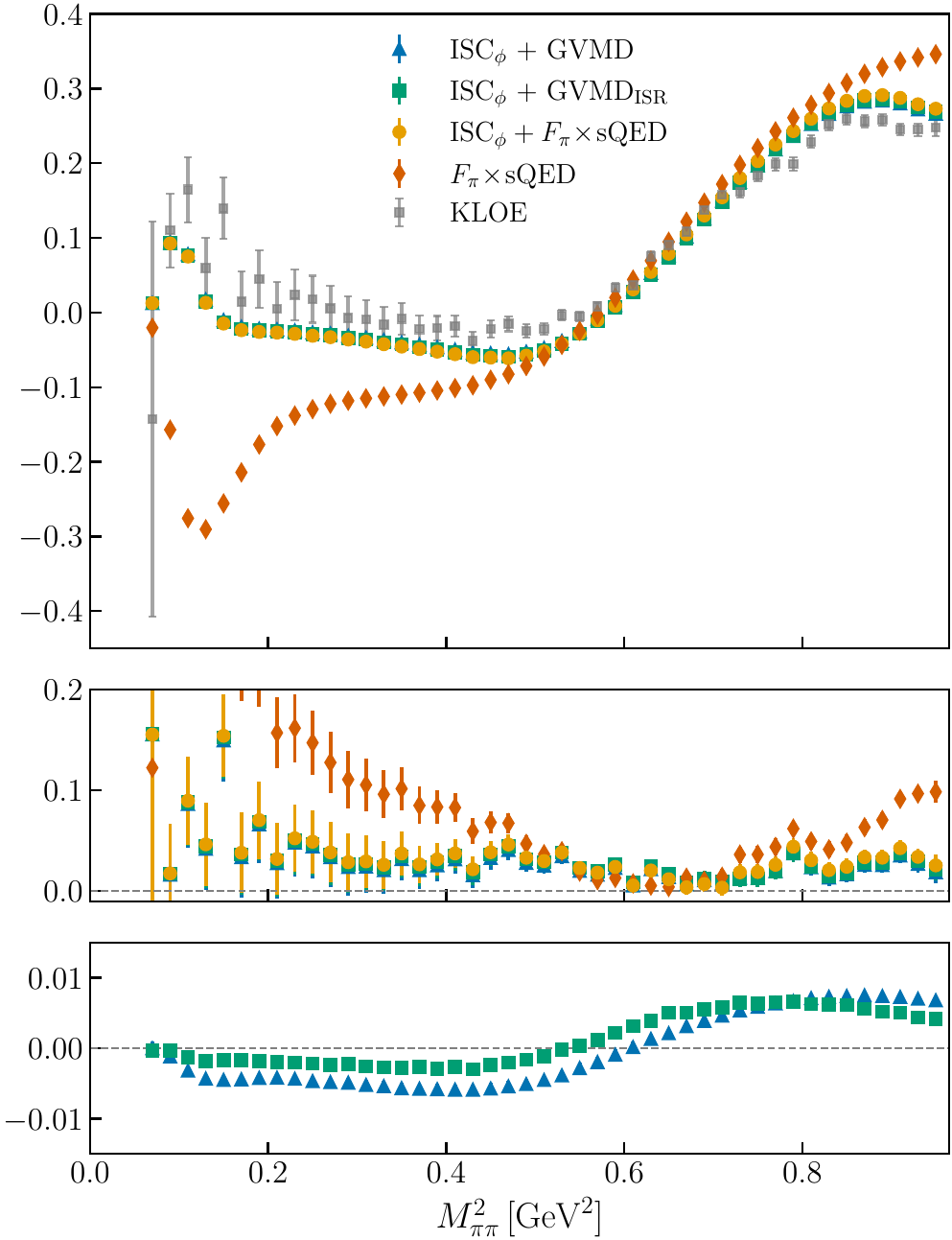}
    \caption{In the top panels we show the forward--backward asymmetry for a relaxed LA KLOE selection at $\sqrt{s}=1~\text{GeV}$ (left) and $\sqrt{s}=1.02~\text{GeV}$ (right). The KLOE experimental data is shown in grey, while the coloured points correspond to \texttt{Phokhara} predictions including different contributions (see text for details). The middle panels display the absolute difference between data and each prediction, $\Delta^{\text{KLOE}}_{X}=\text{KLOE}-X$. The bottom panels compare the MC approaches: for the left plot we take \FxsQED as the reference, while for the right plot we use \FxsQED+ ISC$_\phi$. }
    \label{fig:AFB}
\end{figure*}

In Fig.~\ref{fig:AFB} we show the KLOE data for the forward--backward asymmetry for both centre of mass energies on-peak (right) and off-peak (left) the $\phi$ resonance. On top, we display the different MC contributions, that is, only the NLO \FxsQED corrections, the additional ISC$_\phi$ vector resonances mentioned above, and finally ISC$_\phi$ plus the GVMD$_\text{ISR}$ and full GVMD modifications. On the middle panels we display the difference between data and approaches, and compare the MC choices on the bottom plots. There are several points to highlight:
\begin{enumerate}
    \item The additional ISC$_\phi$ contributions are essential to describe KLOE data at $\sqrt{s} = M_\phi$. They should also be accounted for in scenarios from Subsection~\ref{subsec:kloela}~and~\ref{subsec:kloesa}.  
    \item All MC models deviate from data at high and low $M_{\pi\pi}^2$, while giving better results in the $0.4~\lesssim~ M_{\pi\pi}^2~\lesssim~0.8$ window.
    \item GVMD does not improve the description of the forward--backward asymmetry within the current experimental accuracy. This is expected to be mitigated with new data being analysed in KLOE-nxt. If we restrict ourselves to the region $0.4~<~ M_{\pi\pi}^2~<~0.8$, the mean absolute difference between data and ISC$_\phi$ is $\langle|\Delta_{\text{ISC}_\phi}^\text{KLOE}|\rangle=0.029(2)$, while $\langle|\Delta_{\text{ISC}_\phi+\text{GVMD}}^\text{KLOE}|\rangle=0.025(2)$ for on-peak energies. For $\sqrt{s} = 1$ GeV, $\langle|\Delta_\text{ISC}^\text{KLOE}|\rangle=0.037(1)$, while $\langle|\Delta_\text{ISC+GVMD}^\text{KLOE}|\rangle=0.032(1)$. This points towards a slight improvement by considering GVMD, but the conclusion is highly constrained by the precision of the experimental data. 
    \item In general, differences between MC models are significantly smaller than the difference between data and MC.
\end{enumerate}

Finally, a direct comparison with the energy-scan scenario is not straightforward.
%it is difficult to make a direct comparison with the energy scan scenario. 
In that case, at the $\rho$ peak, $\mathcal{A}_{FB}$ shows a maximal difference between GVMD and \FxsQED~\cite{Ignatov:2022iou,Budassi:2024whw,Colangelo:2022lzg}.
Such behavior is not observed in the radiative-return channel. Instead, the forward–backward asymmetry remains close to zero, a feature predicted by both GVMD and \FxsQED.
%we do not observe this in the radiative return channel. Instead, the forward--backward asymmetry is near zero, and this is predicted by GVMD but also by \FxsQED. 
Overall, the impact of the $\rho$ resonance seems to be tame, and smaller than the effect observed in the energy scan process. A detailed study of the soft photon limit in the amplitudes derived in Eq.~\eqref{eq:amp}, as well as the interplay between GVMD and ISC$_\phi$ corrections, is left for future work. 

\section{Conclusion}
\label{sec:conclusions}

This work presented the first implementation of the radiative return process $e^+e^-\to\pi^+\pi^-\gamma$ at low energies, incorporating an improved description of the non-perturbative pion–photon interaction. The starting point was the point-like approximation of the pion form factor, treated within the factorized \FxsQED framework. Building on this, a parametric representation of the Compton tensor in terms of Breit–Wigner functions was introduced, enabling an automated implementation of the non-perturbative dynamics within an inspired GVMD framework. This implementation was designed to be interfaced with Monte Carlo generators relevant for low-energy studies of hadronic cross sections in $e^+e^-$ collisions.

This GVMD implementation was validated within the Monte Carlo generator \texttt{Phokhara}, where several phenomenologically relevant scenarios were analysed. 
%
%First, 
We revisited the four benchmark setups introduced in Ref.~\cite{Aliberti:2024fpq}: two KLOE-like configurations, a BESIII-like selection of experimental cuts, and a setup intended to mimic $B$-factory conditions. For both KLOE configurations, the GVMD modifications were found to have per-mille or no impact on the total cross section, while percent-level deviations in the angular distributions were observed. Similarly, in the BESIII-like case effects at the level of $0.2\%$ appeared across all distributions. In contrast, no significant deviations were found for the $B$-factory setup. 

As a first step towards validating the descriptive power of GVMD, we have laid the groundwork to pursue the strategy adopted in the energy-scan method, namely to study the forward--backward asymmetry and confront the \FxsQED and GVMD predictions with experimental data. To do so, we have compared the Monte Carlo generator \texttt{Phokhara}, upgraded with the new GVMD implementation and the relevant ISC$_\phi$ meson resonances, with KLOE data, for a broad large-angle scenario and centre of mass energies on and off the $\phi$-resonance peak. While definitive conclusions on the predictive power of GVMD were hard to reach due to the large experimental uncertainties, they will be mitigated in the new forthcoming experimental analysis, KLOE-nxt. 

Finally, we have provided a code to include the GVMD corrections for the two-photon-exchange channels in any MC event generator. We expect the code, together with ongoing efforts to include FsQED and higher order radiative corrections to improve the overall descriptive power of the MC \texttt{Phokhara}.

\newpage 

\acknowledgments

We are indebted to Fedor Ignatov, Thomas Teubner and Graziano Venanzoni for their unwavering support and for many clarifying discussions since the beginning of this project.
We also thank Martin Hoferichter, Fedor Ignatov and Graziano Venanzoni for valuable comments on the manuscript, Stefan M\"uller for providing the experimental data used in Fig.~\ref{fig:AFB}, and Aidan Wright, Thomas Dave and Giovanni Pelliccioli for discussions.
This work is supported by the Leverhulme Trust, LIP-2021-014.

\appendix 
\section{Fortran implementation of TPE diagrams within GVMD}
\label{Appendix}
We briefly describe the content of the Ancillary files provided in~\cite{zenodo}, as well as how to employ the \texttt{Fortran} routines provided therein. In particular, we share two main functions, one for $A^{(1)}_\text{TPE;ISR}$, and one for $A^{(1)}_\text{TPE;FSR}$, both labeled as $\mathcal{X}$\texttt{\_GVMD(ampGVMD,m12,m32,s14,s15,s23,s24,s35)}, with $ \mathcal{X} \in \{\texttt{ISR},\texttt{FSR}\}$. Both routines work in double precision, with all variables being complex. Each of the routines take as inputs the initial and final state squared masses, \texttt{m12} and \texttt{m32}, as well as the kinematic invariants $s_{ij}=(p_i+p_j)^2$, with the momentum configuration described in Section~\ref{SecI}. They output \texttt{ampGVMD}$\mathcal{X}$. We exemplify their use by applying them to the VFF of Ref.~\cite{Ignatov:2022iou}, with the parameters of the VFF $a_v$, and $\Lambda_v = \texttt{mp}v\texttt{2}$, where $v\in\{1,2,3\}$. For each combination of vector mesons, one can respectively get $\texttt{ampMM} = \delta_\text{MM}(\texttt{mp}v\texttt{2},\texttt{mp}w\texttt{2})$ and $\texttt{ampMN} = \delta_\text{MN}(\texttt{mp}v\texttt{2})-\frac{x}{\texttt{mp}v\texttt{2}-x}\delta_\text{NN}$ by calling
\begin{widetext}
\begin{align}
\notag
&\texttt{AmpMM}\mathcal{X}\texttt{(ampMM}\mathcal{X}\texttt{,intsNN}\mathcal{X}\texttt{,intsMN}\mathcal{X}\texttt{,intsMM}\mathcal{X}\texttt{,m12,m32,mpv2,mpw2,s14,s15,s23,s24,s35)}\\
\notag
&\texttt{AmpMN}\mathcal{X}\texttt{(ampMN}\mathcal{X}\texttt{,intsNN}\mathcal{X}\texttt{,intsMN}\mathcal{X}\texttt{,m12,m32,mpv2,s14,s15,s23,s24,s35)}
\end{align}
\end{widetext}
Note that the user has to provide the values of $F_\pi^\dagger(s_{12})$ and $F_\pi^\dagger(s_{35})$ inside these routines.

The subroutines rely on \texttt{IntegralsNN}$\mathcal{X}$, \texttt{IntegralsMN}$\mathcal{X}$ and \texttt{IntegralsMM}$\mathcal{X}$, to calculate the needed vectors of tensor integrals, which have a dedicated cache in \texttt{Collier}. They take as input the same kinematic invariants as the \texttt{Amp} routines. Furthermore, the kinematic coefficients in the amplitudes are calculated by the \texttt{cofMM}$\mathcal{X}$ and \texttt{cofMN}$\mathcal{X}$ routines, generated automatically using \texttt{Form}'s optimizer. 

In total, there are 20 files: two that contain the main routines $\mathcal{X}$\texttt{\_GVMD}, four more, one for each amplitude to be calculated (two origins from the hard photon times two groups: MM and MN), and four extra for each evaluation of kinematic coefficients. An additional six, one per vector of tensor integrals to evaluate (again, FSR and ISR times NN, MN and MM integrals). The four additional files organize the cache and options of \texttt{Collier} (in  \texttt{modInts.f90}), interface the different routines, \texttt{procInterfaces.f90}, compiles the example \texttt{compile.sh} and finally acts as main, \texttt{dFF.f90}.   

\section{The ISC$_\phi$ corrections}\label{Appendix:ISC}

At centre-of-mass energies near $M_\phi$, corrections from including meson decay processes are large and significant~\cite{Shekhovtsova:2009yn, Moussallam:2013una, PhysRevD.81.014010, GALLEGOS2010467}. In particular, beyond the FSR part of the ISC related to the \FxsQED Bremsstrahlung process,
 \begin{equation}\label{eq:FxsQED}
e^+e^-\to\gamma^*\to V \to\pi^+\pi^-\gamma\,,
\end{equation}
the decay via scalar resonance and double resonance processes,
\begin{equation}
\label{eq:phi_decays}
\begin{aligned}
e^+e^-\to\gamma^*\to V \to S\gamma\to\pi^+\pi^-\gamma\,, \\
e^+e^-\to V_1\to(V_2\pi)\gamma\to\pi^+\pi^-\gamma\,,
\end{aligned}
\end{equation}%
are large, and need to be taken into account.
For the considered energy regions, $S(J^{PC} = 0^{++}) = \{f_0, \sigma$\} are the possible intermediate scalar mesons leading to the two-pion final state, while only the lowest nonet of the vector mesons, $V = \{\rho,\omega,\phi\}$, has to be considered. 

To account for the resonances~\eqref{eq:phi_decays}, present in the ISC, the Lorentz tensor structure of the decay process $\gamma^* \to \pi\pi\gamma$, studied in Ref.~\cite{Dubinsky:2004xv}, is elaborated and subsequently incorporated into {\tt Phokhara}, together with the modifications introduced by the GVMD approach.
This was previously implemented in the {\tt FASTERD} Monte Carlo generator~\cite{Shekhovtsova:2009yn}, as well as in an independent version of, \texttt{Phokhara 6.1}. 
It is also worth noting that a related general tensor structure was implemented for the cross-invariant process $\gamma^*\gamma^* \to \pi\pi$ within a dispersive description of the hadronic light-by-light tensor in Refs.~\cite{Hoferichter:2019nlq,Colangelo:2015ama}.

The processes~\eqref{eq:FxsQED} and~\eqref{eq:phi_decays} share the same  tensor structure for the hadronic current. 
In general, the process
$\gamma^* (Q)\to \pi(p_3)\pi(p_5)\gamma(p_4)$, with  $Q=p_1 + p_2$ (in absence of ISR) according to the kinematic configuration described in Section~\ref{SecI}, 
can be expressed in terms of three independent tensor structures:
\begin{align}
\label{eqn:fsr}
M^{\mu \nu}_{\gamma^* \to \pi\pi\gamma} = -ie^{2}\sum_{i=1}^{3} f_i (Q^2, k \cdot Q, k \cdot l)\,\tau_{i}^{\mu \nu }\,,
\end{align}
with 
\begin{align}
\tau _{1}^{\mu\nu}&=k^{\mu }Q^{\nu }-g^{\mu \nu }k\cdot Q\,,
\\
\tau _{2}^{\mu\nu}
&=k\cdot l(l^{\mu }Q^{\nu }-g^{\mu \nu }Q\cdot l)
+l^{\nu }(k^{\mu }Q\cdot l-l^{\mu }k \cdot Q), \;  \nonumber \\
\tau _{3}^{\mu \nu }
&=Q^{2}(g^{\mu \nu }k\cdot l-k^{\mu }l^{\nu})+Q^{\mu }(l^{\nu }k\cdot Q-Q^{\nu }k\cdot l)\,,
\nonumber
\end{align}
$f_i$ Lorentz-invariant functions, so called hadronic structure functions, and $l=p_3-p_5$ and $k=p_4$.
The construction of these independent tensor structures makes use of gauge-invariance constraints, together with the transversality and on-shell conditions of the real photon. Further details can be found in Ref.~\cite{Drechsel:1996ag}.

Because of internal resonances, the total hadronic structure functions are further decomposed as:
\begin{align}
f_{i} = f_i^V + f_i^S + f_i^{Br}\,  , \, \, \, i = 1, 2, 3\,.
\end{align}

Let us emphasise that the decomposition~(\ref{eqn:fsr}) is model independent; the model dependence resides only in the explicit form of the functions $f_i$.
There is a wide variety of models for the $\gamma^*\to S\gamma \to \pi^+\pi^-\gamma$ production mechanism which differ on the treatment of the scalar mesons and their couplings to the virtual photon~\cite{PhysRevD.81.014010, GALLEGOS2010467, Achasov_1998, Isidori_2006, Shekhovtsova:2009yn, Moussallam:2013una}. 
In particular, the double-resonance mechanism has been studied within the vector-meson-dominance (VMD) approach and in the more elaborate framework of resonance chiral perturbation theory (R$\chi$PT) in~\cite{Isidori_2006, Pancheri_2008}.

A detailed comparison of the theoretical model predictions and the KLOE data for the forward–backward charge asymmetry was presented in~\cite{Pancheri_2008, WorkingGrouponRadiativeCorrections:2010bjp, GALLEGOS2010467, PhysRevD.81.014010}. 
In summary, for on- and off-peak $\phi$ data, Bremsstrahlung contributions account for the measurements in the 700–900 MeV region, while the inclusion of double-resonance contributions brings the theoretical predictions into agreement with the data at low pion pair invariant mass.

\bibliographystyle{JHEP}
\bibliography{biblio.bib}

\end{document}